\documentclass[twocolumn]{revtex4-1}

\usepackage{graphics}
\usepackage{amsmath}
\usepackage{amssymb}
\usepackage{amsfonts}

\begin{document}
\title{Pair production in a magnetic and radiation field \\ in a pulsar magnetosphere}
\author{M.~M.~Diachenko}
\email{dyachenko.mikhail@mail.ru}
\author{O.~P.~Novak}
\email{novak-o-p@ukr.net}
\author{R.~I.~Kholodov}
\email{kholodovroman@yahoo.com}

\affiliation{The Institute of Applied Physics of National Academy of Sciences of Ukraine, 
58, Petropavlivska Street, 40000, Sumy, Ukraine}


\begin{abstract}
In the present work one- and two-photon pair production in a subcritical magnetic field have been considered.
Two-photon production has been studied in the resonant case, when the cross section considerably increases compared to the nonresonant case.
While one-photon pair production is considered to be the main mechanism of plasma generation in a pulsar magnetosphere, we suggest the existence of another one, which is resonant two-photon production process.

\vspace{1ex}
\noindent
\textit{Mod. Phys. Lett. A~\textbf{30}, 1550111 (2015);} 
\textbf{doi:} http://dx.doi.org/10.1142/S0217732315501114
\end{abstract}

\maketitle

\section{Introduction}

Astrophysics has a longstanding interest in the physics of elementary processes in strong magnetic field due to the discovery of neutron stars with	field strength of the order of $10^{12}-10^{13}$~G.
Even more remarkable are magnetars with surface fields possibly as high as $10^{15}$~G\cite{Harding06}. 
The fields of these objects are well above the quantum critical magnetic field of $B_c = m^2 c^3 /e\hbar$ ($4.413 \times 10^{13}$~G for electrons), where physical processes are profoundly changed and new processes unique to strong fields dominate. 

The production of electron–positron pairs is an important element in pulsar models, because the presence of an electron–positron plasma is believed to be a necessary condition for the generation of pulsar emission. 
The single-photon pair creation process $\gamma \rightarrow e^- + e^+$ is commonly considered as the main source of pairs, following Ref.~\cite{Sturrock71}.

Pair creation can also result from the interaction between two $\gamma$-ray photons, created through an inverse Compton scattering process by thermal X-ray photons emitted by the surface of the star\cite{Asseo03}.
The attenuation length of a typical gamma-ray for the two-photon process is usually larger than that of one-photon, so that  two-photon pair production has traditionally not been considered important in comparison to one-photon production\cite{Harding06,Burns84,Harding02}.
An exception probably is magnetar supercritical field where 1$\gamma$ production is strongly supperessed by photon splitting\cite{Zhang01, Baring01}.

However, these conclusions concerning 2$\gamma$ process does not take into account the effect of the strong magnetic field near the neutron star surface.
Magnetic field modifies the process, allowing its resonant behavior.
At the resonance the cross section grows significantly and the 2$\gamma$ process, in principle, can make noticeable contribution to plasma generation.
It should be noted that nonresonant two-photon pair production was studied in Refs.~\cite{Kozlenkov86,Dunaev12}. One-photon and two-photon annigilation in a magnetic field were compared for the nonresonant case in Ref.~\cite{Wunner79}.

Thus, the detailed study of two-photon pair production in neutron star magnetosphere and its comparison with one-photon process remains topical.

Let us note possibility to observe QED processes in a strong magnetic field in heavy ion collisions.
Simple estimation shows that moving ions generate magnetic field with magnitude up to $10^{15}$~G at the moment of maximum approach even for collision energy below Coulomb barrier.
The magnetic field effects in the process of pair production in ion collisions have been considered for the first time in Refs.~\cite{Soff81, Rumrich87, Soff88}. 
It was concluded that the effect of magnetic field is rather small. 
However, it was suggested in Ref.~\cite{Fomin98} that collisional magnetic field lifetime can significantly exceed the collision time due to the interaction between the field and the produced pair. 
Since equivalent photon approximation is generally applicable for relativistic heavy ion collisions, the process could be roughly described by 2$\gamma$ pair production in a strong magnetic field\cite{Baur07}.

In the present work one- and two-photon pair production in a subcritical magnetic field have been considered.
Two-photon production has been studied in the resonant case, when the cross section considerably increases compared to the nonresonant case. 
In resonance, one of the photons (a ``hard'' photon) has high enough energy to produce a pair, $\omega \gtrsim 2m$. 
The other photon (a ``resonant'' one) satisfies the resonant conditions and its frequency is a multiple of the cyclotron frequency. 
In this case the intermediate particle become real.

\section{One photon $e^-e^+$ pair production}

It is known\cite{LandauIV} that an electron in a magnetic field occupies discrete energy levels (Landau levels)
\begin{equation}
\label{landau}
  E = \sqrt{p_z^2 + \tilde m^2}, \qquad 
  \tilde{m} = m\sqrt{1 + 2 l b},
\end{equation}
where $p_z$ is parallel to the field momentum, $l$ is the level number and $b$ is magnetic field strength in the units of~$B_c$.

In a magnetic field, energy and parallel momentum conserve,
\begin{equation}
  \label{laws}
    \begin{array}{l}
      E_e + E_p = \omega, \\
      p_{ez} + p_{pz} = k_{z},
    \end{array}
\end{equation}
where $\omega$ is the photon frequency and $k_{z}$ is parallel photon momentum.

The threshold condition following from the conservation laws (\ref{laws}) looks like

\begin{equation}
  \label{threshold}
  \omega^2 - k_{z}^2 = (\tilde m_e + \tilde m_p)^2.
\end{equation}

Apparently, the process is not possible when the photon propagates along the field since in this case $k_z^2 = \omega^2$ and the condition (\ref{threshold}) could not be fulfilled.

Otherwise it is possible to eliminate the parallel photon momentum without loss of generality, $k_z = 0$, since the Lorentz transformation along $\vec B$ does not change the magnetic field.
It can be concluded from (\ref{landau}), (\ref{threshold}) that $p_{ez} =  p_{pz} = 0$ in such reference frame at the threshold.

Hereinafter we assume that $k_z = 0$.
For the photon energy above the threshold the particles momentum looks like
\begin{equation}
\label{pz}
  p_z = \pm \sqrt{ \frac{\omega^2}{4}  -
  m^2  \left[ 1 + b(l_e + l_p)  \right] + 
  \frac{m^2b^2}{\omega^2} (l_e - l_p)^2  }.
\end{equation}


The following electron wave function is used\cite{Fomin00},
\begin{equation}
\label{Psi}
  \Psi_e = \frac{1}{\sqrt{S}}e^{-i(E_et - p_{ey} y - p_{ez} z)} \psi_e(\zeta_e) ,
\end{equation}
where $\zeta_e = m\sqrt{b}(x - p_{ey}/m^2b)$, 
\begin{multline}
\label{psie}
  \psi_e (\zeta_e) = C_e
  \left[
  i\sqrt{\tilde{m}_e - \mu_e m}  U_{l_e}(\zeta_e) + 
  \right. 
  \\   
  \left. 
  + \mu_e\sqrt{\tilde{m}_e + \mu_e m}U_{l_e-1}(\zeta_e)\gamma^1
  \right]u_e .
\end{multline}
Here, $S$ is the normalizing area, $U_l(\zeta_e)$ is the Hermitian function, $\mu_{e}$ is doubled spin projection,  $C_{e}$ is the normalizing constant and $u_{e}$ is the constant bispinor,
\begin{gather}
  \displaystyle
  C_e = \frac 12 \sqrt{\frac{\sqrt{Be}}{E_e\tilde{m}_e}}, \\
  u_{e} = \frac{1}{R_{e}} \left(0,\: - R^2_{e},\: 0,\: p_{e z}\right), \\
  R_{e} = \sqrt{E_{e} - \mu_{e} \tilde m_{e}}.
\end{gather}

\begin{figure}
  \resizebox{\columnwidth}{!}{\includegraphics{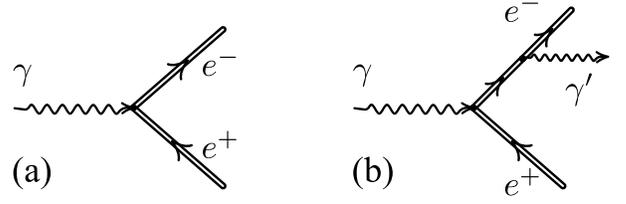}}
  \caption{The Feynman diagrams of (a) 1$\gamma$ pair production and (b) 1$\gamma$ pair production with photon emission. The later rate contains a divergence at $\omega' \rightarrow 0$.}
  \label{pp1g}
\end{figure}

In the first order of perturbation theory the process amplitude of photoproduction is defined by the formula
\begin{equation}
\label{1Sfi}
  S_{fi} = -ie \int d^4x  \:\:
  \bar \Psi_e A^\mu \gamma_\mu \Psi_p,
\end{equation}
where $A^\mu$ is a wave vector of the initial photon\cite{LandauIV}.
The corresponding Feynman diagram is shown in fig.~\ref{pp1g}(a).

We will use the lowest Landau levels approximation,
\begin{equation}
\label{LLL}
  l_{e,p} \sim 1, \quad l_{e,p}b \ll 1.
\end{equation}
This conditions holds true in subcritical field.
In addition, the photon energy is assumed to be close to the threshold value,
\begin{equation}
\label{bias}
    \omega =  \tilde{m}_e + \tilde{m}_p  + \delta \omega,  \qquad \delta \omega \lesssim mb.  
\end{equation}
The expression for the momentum (\ref{pz}) simplifies in this case and reduces to
\begin{equation}
  p_{ez} = \pm \sqrt{m\delta \omega}.
\end{equation}

The process rate reads
\begin{equation}
\label{genrate}
  dW = \frac{|S_{fi}|^2 }{t} \frac{Sd^2p_e}{(2\pi)^2} \frac{Sd^2p_p}{(2\pi)^2},
\end{equation}
where $t$ is time. 
Integration over $d^2p_p$ and $dp_{ez}$ can be carried out using properties of delta-function which appears in the amplitude as a consequence of the conservation laws.
The amplitude does not depend on $p_{ey}$ and integral over $dp_{ey}$ yield an additional factor $p_{ey}$.
To eliminate the nonphysical factor $Sp_{ey}/V$ one should identify normalization length $L = S/V$ with the position of the electron orbit center $x_0 = p_{ey}/m^2b$\cite{Klepikov54}. 
Thus,
\begin{equation}
  \frac{Sp_{ey}}{V} = m^2 b.
\end{equation}

After carrying out corresponding calculations, the process rate
with arbitrary spin projections takes the form\cite{Novak09}:
\begin{gather}
\label{eq81}
W^{++} = \alpha \frac{mb}{2} Y l_e (1 - \xi_3),\\
\label{eq82}
W^{--} = \alpha \frac{mb}{2} Y l_p (1 - \xi_3),\\
\label{eq83}
W^{-+}= \alpha m Y (1 + \xi_3),\\
\label{eq84}
W^{+-} = \alpha \frac{mb^4}{16}
Yl_el_p\left( {({1 + \xi_3 }) + \frac{16\delta\omega }{mb^2}({1 - \xi_3 })}
\right).
\end{gather}
Here, the superscripts denote the electron and positron polarizations respectively, $\xi_3$ is the Stokes parameter defining linear polarization, $l_e$ and $l_p$ are the Landau levels of an electron and a positron,
\begin{gather}
  Y = \frac{be^{ - \eta }}{4 \sqrt{\delta\omega/m}}\frac{\eta ^{l_e + l_p}}{l_e!l_p!},\\
  \eta = \frac{k_x^2 + k_y^2}{2 m^2b} \approx \frac{2}{b}.
\end{gather}

The obtained rates Eqs.~(\ref{eq81})--(\ref{eq84}) contain the denominator $\sqrt{\delta \omega}$ that goes to zero if  the pair produced with zero longitudinal momenta, i.e. at the reaction threshold. 
It results in the occurrence of divergences\cite{Daugherty83}.

Enough attention has been paid to the explanation of the physical nature of these divergences, for example, in Ref.~\cite{Graziani95}, but there is no complete clarity in understanding of this problem.
In our opinion, the presence of singularities is associated with neglected emission of soft photons, which always accompanies quantum-electrodynamics processes. 
This phenomenon is similar to the so-called ``infra-red catastrophe''. 
It is known, that infra-red divergences arise so far as the perturbation theory becomes incorrect for soft photon emission.

The divergence at $\delta \omega \rightarrow 0$ vanishes if an additional final photon is taken into account (see Fig.~\ref{pp1g}(b))\cite{Fomin07}. 
However, the rate of such process is in inverse proportion to the frequency of the final photon, $dW \sim 1 / \omega '$, and diverges if the frequency $\omega'$ goes to zero.

Let us analyze Eqs.~(\ref{eq82})--(\ref{eq84}).
When the pair is produced in the energetically low spin state ($\mu_e = - 1$, $\mu_p =1$) the process has the greatest rate because the corresponding expression (\ref{eq83}) contains the small parameter $b$ in the lowest power. 
In the cases $\mu_e = 1$, $\mu_p = 1$ and $\mu_e = - 1$, $\mu_p = - 1$ the expressions of probability (\ref{eq81}), (\ref{eq82}) differ from Eq.~(\ref{eq83}) in the sign of the parameter of linear polarization $\xi_3$. 
It should be noted that the similar effect takes place in the process of synchrotron radiation. 
Finally, if particles are created in the energetically high spin state with $\mu_e = 1$, $\mu_p = -1$, then the process has the smallest probability.

The process rate in the energetically low spin state Eq.~(\ref{eq83}) vanishes when $\xi_3 = -1$. 
Taking into account the next order in small parameter $b$ the rate $W^{-+}$ takes on the following form,
\begin{equation}
\label{eq106}
 W^{-+} =\alpha m Y  (1 + \xi_3) 
 \left[ 
 1 + \frac{b}{2}\left( 3(l_e + l_p) - 2l_e l_p \right) 
 \right] .
\end{equation}
One can see that dependence on photon polarization remains the same as in the previous case. 
Thus, $W^{-+}$ can be neglected in comparison with $W^{++}$ and $W^{--}$ in the case of perpendicular polarization ($\xi_3 = - 1$).

Let us find the polarization degree of electrons. 
If $\xi_3 \ne - 1$,  the contribution $W^{ - + }$ exceeds all other terms, therefore  $W^{ + + }$ and $W^{ - - }$ can be neglected. 
Consequently,
\begin{equation}
\label{eq109}
P_e \approx - 1.
\end{equation}

To find the more accurate expression of the polarization degree one have to expand $P_e$ in a power series  in the first order in small parameter $b$. 
After simple calculations, the polarization degree takes on the form
\begin{equation}
P_e = - 1 + bl_e \frac{1 - \xi_3 }{1 + \xi_3 }.
\end{equation}

In the case $\xi_3 \to - 1$, the quantity $W^{ - + }$ can be neglected compared to $W^{--}$ and $W^{++}$, thus,
\begin{equation}
\label{eq110}
P_e = \frac{l_e - l_p}{l_e + l_p}.
\end{equation}

The process has the maximal probability when an electron and a positron are produced at the same Landau level, therefore polarization degree converges to zero if the Landau level numbers increase.

\section{Two-photon $e^-e^+$ pair production}

Two photon $e^-e^+$ pair production is the second-order process in the fine structure constant and its amplitude reads 
\begin{multline}
\label{Sfi}
  S_{fi} = -ie^2 \int d^4x d^4x' 
  \left[
  \bar\Psi_e (A_1\gamma) G(x-x') (A'_2\gamma) \Psi_p' +
  \right.
  \\
  \left.
  \bar\Psi_e (A_2\gamma) G(x-x') (A'_1\gamma) \Psi_p'
  \right].
\end{multline}

\begin{figure}
  \resizebox{\columnwidth}{!}{\includegraphics{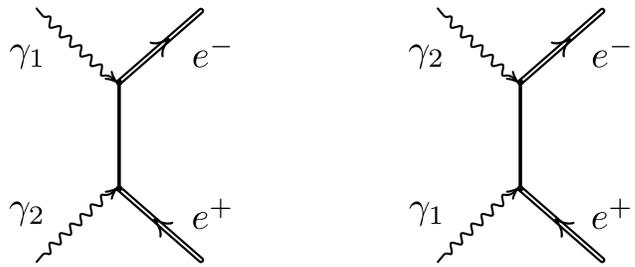}}
  \caption{Feynman diagrams of two-photon $e^-e^+$ pair production in a magnetic field.}
  \label{fig2}
\end{figure}

Figure~\ref{fig2} shows the Feynman diagrams that corresponds to the amplitude (\ref{Sfi}). 

The virtual electron propagator has the form\cite{Fomin99}:
\begin{equation}
\label{G}
  G(x-x') = \frac{-m\sqrt{b}}{(2\pi)^3} 
  \int dg_0 dg_y dg_z e^{-i\Phi} \sum\limits_{n}   \frac{G_H(x,x')}{g_0^2 - E_n^2},
\end{equation}
\begin{multline}
\label{GH}
  G_H(x,x') = U_n U_n' (\gamma P + m) \tau + im \sqrt{2nb} U_{n-1}U'_{n}\gamma^1\tau - \\
  - im \sqrt{2nh} U_n U_{n-1}' \tau \gamma^1 + U_{n-1} U'_{n-1} (\gamma P + m)\tau^* ,
\end{multline}
where $\gamma$ are the Dirac gamma matrices,
\begin{gather}
  \Phi = g_0(t-t') - g_y (y-y') - g_z (z-z'), \\
  \tau = \frac{1}{2} (1 + i \gamma_2\gamma_1), \\
  P = (E_n, 0, 0, g_z).
\end{gather}
Hermitian functions $U_n$ depend on the argument
\begin{equation}
  \rho(x) = m\sqrt{b} x + g_y/m\sqrt{b},
\end{equation}
and primed functions depend on $\rho(x')$.

A resonance occurs when kinematics allows virtual electron to be on shell and quantities $g_0$, $g_z$ satisfy the relation between electron energy and momentum in a magnetic field,
\begin{equation}
\label{rescond}
  g_0^2 = g_z^2 + m^2 + 2nbm^2 = E_n^2.
\end{equation}

In this case the denominator in the propagator (\ref{G}) goes to zero.
To eliminate the divergence one should introduce the radiative width $\Gamma$ according to the Breit-Wigner prescription\cite{Graziani95},
\begin{equation}
  \label{Breit}
  E_n \rightarrow E_n - \frac{i}{2} \Gamma.
\end{equation}

The process kinematics is determined by the same conservation laws Eq.~(\ref{laws}) with replacements $\omega \rightarrow \omega_1 + \omega_2$ and $k_z \rightarrow k_{1z} + k_{2z}$.
Note that Lotentz transformation to a frame moving along magnetic field does not change the field $\vec{H}$ itself.
Thus, $z$-component of the total photon momentum can be eliminated by the choice of the reference frame,
\begin{equation}
\label{perp}
  k_{1z} + k_{2z} = 0.
\end{equation}
Taking into account the conservation laws in each vertex and condition (\ref{rescond}), it is easy to obtain the following expressions for resonant photon frequencies,
\begin{equation}
\label{resfreq}
  \begin{array}{l}
    \omega_1 = mb (l_e - n), \\
    \omega_2 = 2m + mb(l_p + n) + \delta \omega.
  \end{array}  
\end{equation}
Apparently, the pair is created by the hard photon $\omega_2$ while the resonant one $\omega_1$ induces the transition of the intermediate electron between energy levels.
This is not an unexpected result since resonant process can be viewed as consequent one-photon pair production and photon absorption.

Let us estimate resonant cross-section of two-photon pairproduction.
The lowest possible level numbers are chosen to fulfill the conditions (\ref{bias})--(\ref{resfreq}) and the energetically favourable spin state is chosen for the produced pair, 
\begin{gather}
\label{lowres}
  l_e = 1, \qquad l_p = n =0, \\
\label{favor}
  \mu_e = -1, \qquad \mu_p = +1.
\end{gather}

In Eq.~(\ref{G}) all summands can be neglected except the first one with $n = 0$, where the denominator goes to zero.
The second diagram can be neglected as well, since its resonant conditions can not be satisfied with level numbers defined by Eq.~(\ref{lowres}).

Cross-section is defined as process rate divided by flux~$j$,
\begin{gather}
\label{gencross}
  d\sigma = \frac{|S_{fi}|^2 }{t \: j} \frac{Sd^2p_e}{(2\pi)^2} \frac{Sd^2p_p}{(2\pi)^2},\\
  j = (1 - \cos \chi)/V,
\end{gather}
where $\chi$ is the angle between the initial photons.

After developing the amplitude into a series in $b$ and carrying out the integrations over the interval of the final states, the cross-section takes on the form
\begin{equation}
  \label{final}
  \sigma =  \frac{\alpha^2\pi b e^{-\frac{2}{b}}}{8(1 - \cos\chi)}
  \sqrt{\frac{m}{\delta \omega}}
  \frac{(1 + \xi_3^{(2)})(1+u^2+2u\xi_2^{(1)}- s^2\xi_3^{(1)})}{ (g_0 - E_n)^2  + \frac{\Gamma^2}{4} },
\end{equation}
where $u = \cos\theta$, $s = \sin\theta$ and $\theta$ is the polar angle of the resonant photon.

\section{Conclusion}

In the present work one and two photon pair production in  a subcritical magnetic field have been considered.
These processes are believed to be the mechanisms of plasma generation in a pulsar magnetosphere.
2$\gamma$ production has been studied in the resonant case. 
The resonant cross section considerably increases compared to the nonresonant case, so that 2$\gamma$ production can make noticeable contribution to plasma generation in a magnetosphere.

It is convenient to illustrate the relation between 1$\gamma$ and 2$\gamma$ production considering another process, double synchrotron emission.
The differential rate of double emission is given by
\begin{equation}
  \frac{dW_{2\gamma}}{d\omega_1 du dv} = \frac{1}{4\pi} 
  \frac{\frac{dW_{l, n}}{du}\frac{dW_{n, l'}}{dv}}{(\omega_1 - \omega_{1res})^2 + \Gamma^2/4}.
\end{equation}
Integrating of this expression over $\omega$ at the resonance is trivial and results in additional factor $\pi\Gamma/4$. 
Taking into account that $\Gamma \sim W_{rad}$ and carrying out integration over $du,dv$ one obtains the following relation on the rates,
\begin{equation}
  W_{2\gamma} \sim W_{1\gamma}.
\end{equation}

It is possible to estimate the pair production rate when a hard photon with a frequency $\omega = 2m + \delta\omega$ propagates in a magnetosphere with magnetic field $b$ and photon density $n_\gamma$. 
For the sake of simplicity we will asume all the magnetosphere photons to have the synchrotron frequency $\omega = mb$ and, thus, satisfy the resonant conditions for the two-photon process.
The corresponding rate reads
\begin{equation}
\label{nu2}
  \nu_{2\gamma} = n_\gamma \sigma (1 - \cos \chi),
\end{equation}
where $\sigma$ is defined by Eq.~(\ref{final}) with radiative width
\begin{equation}
  \Gamma \approx \frac{2}{3}\alpha m b^2.
\end{equation}
Note that synchrotron photon density $n_\gamma$ enters Eq.~(\ref{nu2}) as a free parameter. 

The rate of the one-photon process is\cite{Novak09}
\begin{equation}
  \nu_{1\gamma} \approx \frac{\alpha mb}{4\sqrt{\delta \omega /m}} e^{-\frac{2}{b}}.
\end{equation}
From the above expressions follows, that the two photon production dominates when photon density exceedes the critical value
\begin{equation}
\label{nc}
  n_{\gamma c} = \frac{2}{9\pi}\frac{\alpha b^4}{\tilde{\lambda}_c^3},
\end{equation}
where   $\tilde{\lambda}_c$ is Compton wavelength.
For example, when magnetic field strength is $b=0.1$ then the critical density is $n_{\gamma c} \sim 10^{25}$~cm$^{-3}$.

Thus, we would like to point out the existence of another competing mechanisms of plasma generation in a pulsar magnetosphere in addition to the commonly considered one photon pair production.



\begin{thebibliography}{90}

\bibitem{Harding06}
A.~K.~Harding and D.~Lai. \textit{Rep. Prog. Phys.}~\textbf{69} (2006) 2631. 

\bibitem{Sturrock71}
P.~A.~Sturrock. \textit{Astrophys.~J.}~\textbf{164} (1971) 529.

\bibitem{Asseo03}
E.~Asseo. \textit{Plasma Phys. Control. Fusion}~\textbf{45} (2003) 853.

\bibitem{Burns84}
M.~L.~Burns and A.~K.~Harding. \textit{Astrophys.~J.}~\textbf{285} (1984) 747.

\bibitem{Harding02}
A.~K.~Harding, A.~G.~Muslimov and B.~Zhang. \textit{Astrophys.~J.}~\textbf{576} (2002) 366.

\bibitem{Zhang01}
B.~Zhang. \textit{Astrophys.~J.}~\textbf{562} (2001) L59.

\bibitem{Baring01}
M.~G.~Baring and A.~K.~Harding. \textit{Astrophys. J}~\textbf{547} (2001) 929.

\bibitem{Kozlenkov86}
A.~A.~Kozlenkov and I.~G.~Mitrofanov. \textit{Sov. Phys.~JETP}~\textbf{64} (1986) 1173.

\bibitem{Dunaev12}
M.~A.~Dunaev and N.~V.~Mikheev. \textit{JETP}~\textbf{114} (2012) 365.

\bibitem{Wunner79}
G.~Wunner. \textit{Phys. Rev. Lett.}~\textbf{42} (1979) 79.

\bibitem{Soff81} 
G.~Soff, J.~Reinhardt and W. Greiner. \textit{Phys. Rev.~A}~\textbf{23} (1981) 701.

\bibitem{Rumrich87}
K.~Rumrich, W.~Greiner and G.~Soff. \textit{Phys. Lett.~A}~\textbf{125} (1987) 394.

\bibitem{Soff88}
G.~Soff, J.~Reinhardt. \textit{Phys. Lett.~B}~\textbf{211} (1988) 179.

\bibitem{Fomin98}
P.~I.~Fomin, R.~I.~Kholodov. \textit{Reports of NAS of Ukraine}~\textbf{12} (1998) 91.

\bibitem{Baur07} 
G.~Baur, K.~Hencken and D.~Trautmann. \textit{Phys. Rep.}~\textbf{453} (2007) 1.

\bibitem{LandauIV}
V.~B.~Berestetskii, E.~M.~Lifshitz, and L.~P.~Pitaevskii, \textit{Relativistic Quantum Theory} (Pergamon Press, Oxford,1982).

\bibitem{Fomin00}
P.~I.~Fomin and R.~I.~Kholodov, \textit{Zh. Exp. Theor. Phys.}~\textbf{117} (2000) 319.

\bibitem{Klepikov54}
N.~P.~Klepikov, \textit{Zh. Exp. Theor. Phys.}~\textbf{26} (1954) 19.

\bibitem{Novak09}
O.~P.~Novak and R.~I.~Kholodov. \textit{Phys. Rev.~D}~\textbf{80} (2009) 025025.

\bibitem{Daugherty83}
J.~K.~Daugherty and A.~K.~Harding. \textit{Astrophys.~J.}~\textbf{273} (1983) 761.

\bibitem{Graziani95}
C.~Graziani, A.~K.~Harding and R.~Sina, \textit{Phys. Rev.~D}~\textbf{51} (1995) 7097.

\bibitem{Fomin07}
P.~I.~Fomin and R.~I.~Kholodov, \textit{Probl. At. Sci. Tech.}~\textbf{3} (2007) 179.

\bibitem{Fomin99}
P.~I.~Fomin and R.~I.~Kholodov, \textit{Ukr. phys.~j.}~\textbf{44} (1999) 1526.

\end{thebibliography}
\end{document}